\documentclass[letterpaper, 10 pt, conference]{ieeeconf}  

\IEEEoverridecommandlockouts                              
\let\proof\relax

\usepackage{dsfont}
\usepackage{amsthm}
\usepackage{amsmath}
\usepackage{amssymb,amsfonts}
\usepackage{dsfont}
\usepackage{bbm}
\usepackage{paralist}
\usepackage{xcolor}
\usepackage{mathtools}
\usepackage{booktabs}
\usepackage{cite}
\usepackage{url}
\usepackage[linesnumbered,ruled,vlined]{algorithm2e}

\newtheorem{theorem}{Theorem}
\newtheorem{assumption}{Assumption}

\newtheorem{lemma}{Lemma}

\newtheorem{remark}{Remark}

\def\NN{{\mathbb N}}

\def\RR{{\mathbb R}}

\newcommand{\ep}{\hfill $\Box$}

\newcommand{\set}[1]{\mathcal{#1}}
\newcommand{\norm}[1]{\lVert#1\rVert}

\title{\LARGE \bf
Self-Tuning Tube-based Model Predictive Control
}

\author{Damianos Tranos, Alessio Russo, and Alexandre Proutiere
\thanks{This work was supported by the Wallenberg AI, Autonomous Systems and Software Program (WASP) funded by the Knut and Alice Wallenberg Foundation.}
\thanks{D. Tranos, A. Russo, and A. Proutiere are with the Division of Decision and Control Systems, School of Electrical Engineering and Computer Science, Royal Institute of Technology (KTH), Stockholm, Sweden. Emails: 
        \{{\it tranos@kth.se, alessior@kth.se, alepro@kth.se }\}.}%
}

\begin{document}

\maketitle
\begin{abstract}
We present Self-Tuning Tube-based Model Predictive Control (STT-MPC), an adaptive robust control algorithm for uncertain linear systems with additive disturbances based on the least-squares estimator and polytopic tubes. Our algorithm leverages concentration results to bound the system uncertainty set with prescribed confidence, and guarantees robust constraint satisfaction for this set, along with recursive feasibility and input-to-state stability. Persistence of excitation is ensured without compromising the algorithm's asymptotic performance or increasing its computational complexity. We demonstrate the performance of our algorithm using numerical experiments.
\end{abstract}

\section{Introduction}

Model Predictive Control (MPC) \cite {mayne2000constrained} addresses the infinite horizon optimal control problem in the presence of input and state constraints by approximating it as a sequence of finite horizon optimization problems. When the dynamics of the system are uncertain, robust MPC methods \cite{mayne2005robust} can be employed to ensure constraint satisfaction for pre-specified sets of system parameters and disturbances. This robustness comes at the cost of a reduced closed-loop performance. To mitigate this performance loss, one may leverage adaptive control techniques \cite{mayne2014model} to learn the system dynamics in an online manner, and in turn reduce the uncertainty causing this loss. 

Adaptive control schemes have been developed and studied mainly for unconstrained control problems. The first results \cite{aastrom1973self} concerned their asymptotic convergence properties and established conditions under which the controller derived from these schemes actually approaches the optimal feedback controller obtained by assuming the full knowledge on the system dynamics. More recently, see e.g. \cite{abbasi2011regret,mania2019certainty,jedra2022minimal}, researchers managed to quantify the convergence rate of some classical adaptive control schemes, such the celebrated self-tuning regulators, as well as the price that one has to pay (in terms of cumulative losses -- often captured through the notion of regret) to learn the system dynamics. These recent important results are however restricted to the LQR problems in unconstrained linear systems.

In this paper, we investigate the design and the performance analysis of MPC schemes handling system dynamics uncertainties. We propose to combine adaptive and robust control methods. The adaptive control component of our schemes allows to rapidly reduce over time the uncertainties in a controlled and quantifiable manner, whereas the robust control component ensures constraint satisfaction. More precisely, our contributions are as follows.

\medskip
\noindent
{\bf Contributions.} We address the problem of controlling an uncertain linear system with parametric and additive disturbances, subject to deterministic constraints. We present Self-Tuning Tube-based Model Predictive Control (STT-MPC), an algorithm combining adaptive and robust control techniques. STT-MPC uses a simple Least Squares Estimator (LSE) for estimating the dynamics and a parameter set compatible with the observations with prescribed level of certainty. We derive tight concentration results (similar to those presented in \cite{jedra2022minimal} for the LQR problem) for this set, and exploit these results to construct fixed complexity polyhedral approximations of it. In STT-MPC, these approximations are used to build a polytopic tube MPC scheme \cite{fleming2014robust} to ensure robust constraint satisfaction. We establish the recursive feasibility and the input-to-state stability of the proposed scheme.

In contrast to previously proposed robust MPC schemes (refer to \textsection\ref{sec:related_work} for details), STT-MPC enjoys the following properties. (i) STT-MPC uses a probabilistic rather than robust estimation scheme, leading to notably faster adaptation rates and smaller parameter sets for which the constraints need to be satisfied. (ii) Persistence of Excitation (PE), required to get performance guarantees for the LSE, is achieved without modifying either the cost or objective function of the MPC. The added excitation is treated as an additive disturbance and can be chosen to decay to zero with time (e.g., at a rate $1/\sqrt{t}$). This allows us to asymptotically recover the performance of the standard MPC algorithm with full knowledge of the dynamics. (iii) An input-to-state stability analysis of STT-MPC is possible even when the PE condition is active and without imposing any additional restrictions on the choice of the estimate of the nominal model parameter. 

\medskip
\noindent
{\bf Notations.} For a time dependent vector $x_t$, we denote by $x_{k|t}$ its prediction at time $k+t$ given information at time $t$. For any two sets $\set{A}$ and $\set{B}$, we define their Minkowski sum as the set $\set{A}\oplus\set{B} := \{a+b: a\in \set{A}, \ b\in \set{B}\}$. We also define, for any constant $\lambda \geq 0$, the scaled set $\lambda \set{A} := \{\lambda a, \ a \in \set{A}\}.$ For any $d \in \NN$, $x \in \RR^d$, and $\epsilon > 0$ let $\set{B}(x,\epsilon)$ denote the $\epsilon$ ball of the spectral norm centered on $x$. The unit ball centered at the origin is denoted simply as $\set{B}$. For any set $\set{S}$, and any $\varepsilon > 0$ there exists a polytope $\set{P}$ that is an outer approximation of $S$, i.e. $\set{S} \subset \set{P} \oplus \varepsilon \set{B}$. We refer to this polytope as the outer polyhedral approximation of $\set{S}$. Finally, a function $\kappa: \RR_+ \to \RR_+$ is of class $\set{K}$ if it is strictly increasing and $\kappa(0) = 0$ and is of class $\set{K}_\infty$ if in addition $\kappa(x) \to \infty$ as $x \to \infty$.

\section{Related Work}\label{sec:related_work}

An essential ingredient of adaptive control is Persistent Excitation (PE) \cite{narendra1987persistent} (which is equivalent to the notion of required exploration in reinforcement learning \cite{tranos2021regret}). In the context of MPC, different solutions for PE have been investigated:  \cite{marafioti2014persistently} introduces a sufficient PE condition as an additional constraint in their MPC optimization. On the other hand, \cite{heirung2015mpc}, \cite{heirung2017dual}, and later \cite{cai2021dual}, propose dual-MPC schemes where the predicted parameter covariance matrix is included in the cost function, though these dual-control methods lack feasibility and stability guarantees.

In a different direction, the works of \cite{aswani2013provably} and \cite{di2016indirect} introduce robust MPC schemes with guaranteed recursive feasibility and constraint satisfaction. These schemes come with the cost of only considering adaptation for the nominal model, where PE is not ensured, and robustness is guaranteed for a fixed parameter set that is not updated online. In contrast, our scheme, STT-MPC, updates the parameter set each step, and the latter rapidly concentrates around the true paraemter. 

In \cite{tanaskovic2014adaptive}, the authors combine set-membership identification, which involves updating a set of parameters compatible with the observed state-trajectory, and robust constraint tightening for FIR models. This approach was later extended in \cite{lorenzen2019robust} to the case of linear state-space models where online set-membership is combined with homothetic tube MPC \cite{rakovic2012homothetic}. Subsequently, \cite{lu2019robust} proposed an adaptive robust MPC scheme using again set-membership identification and a less conservative polytopic tube MPC scheme \cite{fleming2014robust}. As in \cite{marafioti2014persistently}, they ensure PE in the form of added convex constraints and a modification of the cost function.

\section{Model, Objective, and Approach}

\subsection{Model and assumptions}

We consider the following discrete time, linear, time-invariant system:
\begin{align}
x_{t+1} =  A(\theta^\star) x_t + B(\theta^\star) u_t + w_t,
\label{eq:linear_system}
\end{align}
where $x_t, w_t\in \mathbb{R}^{d_x}$ and $u_t\in \mathbb{R}^{d_u}$. The state transition and state-action transition matrices $A(\theta^\star)$ and $B(\theta^\star)$ are initially unknown. The set of possible such matrices is parametrized by $\theta\in \mathbb{R}^{d_\theta}$ (here $\theta$ could well parametrize each entry of the matrices, in which case $d_\theta=d_x(d_x+d_u)$). To simplify the notations, for two possible parameters $\theta_1,\theta_2$, we define $\norm{\theta_1 - \theta_2}:= \max(\norm{A(\theta_1) - A(\theta_2)}_2,\norm{B(\theta_1) - B(\theta_2)}_2)$. We make the following assumptions.
\begin{assumption}[Parameter uncertainty]
For some $\epsilon_0>0$, ${\cal B}(\theta^*,\epsilon_0)\subset \Theta_0$ where $\Theta_0$ is a known convex polytope.
\label{ass:parameter_set}
\end{assumption}

\begin{assumption}[Additive disturbance]
The sequence $(w_t)_{t\ge 0}$ is i.i.d, and for each $t\ge 0$, $w_t$ is zero-mean, isotropic, with support in the ball $\set{B}(0,3\sigma)$. Hence, $w_t$ is $\sigma^2$-sub-gaussian. Further define $\set{W}$, a convex polytope providing a conservative approximation of $\set{B}(0,3\sigma)$, i.e., $\set{B}(0,3\sigma) \subset \set{W}$.
\label{ass:noise}
\end{assumption}

The system also needs to obey the following state and input constraints for all $t \geq 0$,
\begin{align}
F x_t + G u_t \leq \textbf{1},
\label{eq:constraints}
\end{align}
where $F \in \RR^{d_c \times d_x}$ and $G \in \RR^{d_c \times d_u}$ define the state and input constraints respectively. 
\begin{assumption}[State and input constraints]
The set
\begin{align*}
\set{C} = \{(x,u) \in \RR^{d_x}\times \RR^{d_u}:F x + G u \leq \textbf{1}.\}
\end{align*}
is compact and contains the origin in its interior.
\label{ass:constraints}
\end{assumption}

%
Finally, we assume that we have access to a stabilizer $K$:
\begin{assumption}[Stabilizing Controller]
There exists a known, robustly stabilizing feedback gain $K$ such that $A(\theta)+B(\theta)K$ is stable (i.e., $\rho(A(\theta)+B(\theta)K) <1$) for all $\theta \in \Theta_0$.
\label{ass:stabilizing_controller}
\end{assumption}

\subsection{Objective and MPC}

We wish to minimize the long-term cost defined as $\lim\sup_{T\to\infty} \frac{1}{T}\sum_{t=0}^{T-1} x_t^\top Qx_t+u_t^\top Ru_t$, through some positive semi-definite matrices $Q, R$. To this aim, we use MPC, with a receding horizon $N$. Specifically, at time $t$, given the current system state $x_t$ and the past observations used to derive an estimator $\theta_t$ of $\theta^\star$, we will identify a control policy $(u_{k|t})_{k=0,\ldots,N-1}$ minimizing the cost along a predicted system trajectory $(x_{k|t})_{k=0,\ldots,N}$. We use the well-known dual mode prediction paradigm \cite{kouvaritakis2016model} with the following predicted control sequence,
\begin{align}\label{eq:prediction_control}
u_{k|t} = \begin{cases}Kx_{k|t} + v_{k|t} \quad \forall k \in \{0,\dots,N-1 \}, \\ Kx_{k|t} \quad \forall k \geq N, \end{cases}
\end{align} 
where $\{v_{0|t},\dots v_{N-1|t}\}$ are the optimization variables to be determined by the MPC. The resulting prediction dynamics will be for $k \in \{0,\dots,N-1\}$,
\begin{subequations}
\begin{align}
x_{0|t} &= x_t, \\
x_{k+1|t} &= \Phi(\theta_t)x_{k|t} + B(\theta_t)v_{k|t},
\end{align}
\label{eq:prediction_dynamics}
\end{subequations}
where $\Phi(\theta):=A(\theta)+B(\theta)K$ for any $\theta\in \Theta_0$.

\subsection{General approach}

To handle the uncertainty due to both the noise and the fact that $\theta^\star$ is unknown, we apply a tube-based MPC approach. The tube used in step $t$ is essentially constructed from a polytope approximating a ball centered at the LSE $\hat{\theta}_t$ of $\theta^\star$ and whose radius corresponds to a prescribed level of confidence $\delta$ that the user wishes to guarantee. Persistent excitation is achieved by adding (bounded) noise to the input. The resulting algorithm is presented in Algorithm \ref{alg:main}, and its ingredients are detailed in the next section. Its analysis is given in Section \ref{sec:analysis}. 

\begin{algorithm}[ht]
\DontPrintSemicolon
\KwIn{Initial state $x_0$; confidence $\delta$; estimate $\theta_0$; uncertain parameter set $\Theta_0$\;}
Find stabilizing matrix $K$ for all $\theta\in \Theta_0$\;
Compute matrices $T$ and $H_c$ using \eqref{eq:H_c} \;
\For{$t=1,\dots,\set{T}$}{
\If{$t < t^\star(\delta)$}{
Set $\theta_t \gets \theta_{t-1}$ and  $\Theta_t \gets \Theta_{t-1}$ \;
 }
 \Else{
Update $\theta_t\gets\hat{\theta}_t$ using \eqref{eq:LSE} \;
Compute $\Delta_t$ and set $\Theta_t \gets \Theta_{t-1} \cap \Delta_t$\;
 }
Compute matrices $H_t^{(j)}$ using \eqref{eq:H^j} \;
Obtain $P(\theta_t)$ by solving \eqref{eq:lyapunov} \;
Solve $\set{P}_N(x_t,\theta_{\rho(t)},\Theta_{\rho(t)},\bar{\set{W}}_t)$ \;
Apply $u_t = Kx_t + v^*_{0|t} + \zeta_t$\;
} 
\caption{STT-MPC}
\label{alg:main}
\end{algorithm}

\section{Self-tuning Tube-based MPC}\label{sec:lse}

\subsection{LSE and persistent excitation}

Our algorithm starts with an initial parameter $\theta_0\in \Theta_0$, which is then updated using the LSE. We let $\theta_1=\theta_0$. For $t\ge 2$, the LSE enjoys the following explicit expression:
\begin{align}
\hat{\theta}_t = {\left( \sum\limits_{k=0}^{t-2}x_{k+1} \begin{bmatrix}
x_k \\ u_k
\end{bmatrix}^\top
 \right) \left( \sum\limits_{k=0}^{t-2}\begin{bmatrix}
 x_k \\ u_k
 \end{bmatrix}
 \begin{bmatrix}
 x_k \\ u_k
 \end{bmatrix}^\top
 \right)^\dagger}.
 \label{eq:LSE}
\end{align}
For $t\ge t^\star(\delta)$, we align our prediction parameter $\theta_t$ to $\hat{\theta}_t$. As shown in \cite{jedra2022minimal}, a finite-time analysis of the performance of the LSE is rather intricate but possible even if the feedback controller varies over time. The performance is tightly related to the minimal eigenvalue of the cumulative covariate matrix $\lambda_{\min}(\sum_{s=0}^{t-2}y_sy_s^\top)$ where $y_s=\begin{bmatrix}
x_s \\ u_s
\end{bmatrix}$. More precisely, for the LSE to lead to a good approximation of $\theta^\star$, we need to ensure that this eigenvalue grows with time. To this aim, we add an isotropic and bounded noise to the control input. This noise is represented by the random vector $\zeta_t$ taken to be the projection of $\xi_t$ on $\set{B}(0,3\sigma_t)$, where $\xi_t$ is i.i.d. according to a normal distribution, i.e., $\xi_t \sim \set{N}(0,\sigma_t^2 I_{d_u})$. The choice of $\sigma_t$ directly impacts the performance of the LSE but also the overall performance of the controller (a higher $\sigma_t$ means higher excitation and hence better LSE, but at the expense of a worse overall controller, refer to Section \ref{sec:conclusion}). Here we set $\sigma_t^2 = \sqrt{d_x}\sigma^2 t^{-\alpha}$ for some $\alpha\in (0,1)$. This ensures that (i) the LSE $\theta_t$ converges to $\theta^\star$ (refer to Section \ref{sec:analysis} for a precise statement) and (ii) the controller converges to that obtained through a classical tube-based MPC framework with known $\theta^\star$. When analyzing the performance of the LSE, we will show that the {\it good} event ${\cal G}$ holds with probability at least $1-\delta$:
$$
{\cal G}=\left(\| \hat{\theta}_t-\theta^\star\|\le \epsilon_t, \ \forall t\ge t^\star(\delta)\right),
$$
where $t^\star(\delta)=c_1+c_2\log(1/\delta)$ and $\epsilon_t^2 = c_3 \log(t/\delta)/t^{1-\alpha}$ for some positive constants $c_1, c_2, c_3$. From the above result, we define $\Delta _t$ as an outer polyhedral approximation of $\set{B}(\theta_t,2\epsilon_t)$. We further recursively define the {\it uncertainty sets} as follows: $\Theta_t=\Theta_{t-1}\cap \Delta_t$ for all $t\ge t^\star(\delta)$ and $\Theta_t=\Theta_0$ for $t<t^\star(\delta)$. By construction, the true parameter $\theta^\star$ belongs to the interior of $\Theta_t$ with high probability in the following sense:

\begin{lemma}
Under event ${\cal G}$, $\set{B}(\theta^\star,\epsilon_t) \subset \Theta_t$ for all $t\ge 1$.
\label{prop:inclusion}
\end{lemma}

\proof For $t< t^\star(\delta)$ we have $\Theta_t = \Theta_0$ and the result holds by Assumption 1. Let $t\ge t^\star(\delta)$. We show that $\set{B}(\theta^*,\epsilon_t) \subset \Delta_t$ and $\set{B}(\theta^*,\epsilon_t) \subset \Theta_{t-1}$. For the first, we have for all $\theta \in \set{B}(\theta^*,\epsilon_t)$, using the triangle inequality: $\norm{\theta-\theta_t} \leq \norm{\theta-\theta^*} + \norm{\theta_t-\theta^*} \leq 2\epsilon_t$, where the second inequality holds under ${\cal G}$. Thus $\theta \in \set{B}(\theta_t,2\epsilon_t)$ and so $\set{B}(\theta^*,\epsilon_t) \subset \set{B}(\theta_t,2\epsilon_t) \subset \Delta_t$. We prove $\set{B}(\theta^*,\epsilon_t) \subset \Theta_{t-1}$ by induction. Assume that $\set{B}(\theta^*,\epsilon_t) \subset \Theta_{t-1}$. Then we show that $\set{B}(\theta^*,\epsilon_{t+1}) \subset \Theta_{t}$. Let $\theta$ such that $\|\theta-\theta^\star\|\le \epsilon_{t+1}$. Then $\|\theta-\theta_t\|\le \norm{\theta-\theta^*} + \norm{\theta_t-\theta^*}\le \epsilon_{t+1}+\epsilon_t$. This implies that $\theta\in \Delta_t$. In addition, $\theta\in \Theta_{t-1}$. Indeed, $\|\theta-\theta^\star\|\le \epsilon_{t+1}\le \epsilon_t$, and we conclude using the induction assumption. Hence $\theta\in \Theta_t$.\ep

%





\subsection{Polytopic tubes and associated linear constraints}

With the considered control inputs, the system dynamics becomes:
\begin{equation}\label{eq:newdyn}
x_{t+1}=\Phi(\theta^\star)x_t+B(\theta^\star)(v_{0|t}+\zeta_t)+w_t.
\end{equation}
We apply a tube-based approach, and at time $t$, we build a polytopic tube based on:
\begin{itemize}
    \item[(i)] $\Theta_t$, encoding the uncertainty about $\theta^\star$. We denote by $m_t$ the number of vertices of $\Theta_t$, and the vertices themselves by $\theta_t^{(j)}$, $j=1,\ldots,m_t$.
    \item[(ii)] A polytope $\bar{\set{W}}_t$, handling the uncertainty due to the {\it noise} $B(\theta^\star)\zeta_t+w_t$, including that due to the persistent excitation. To define $\bar{\set{W}}_t$, let $\bar{B}_t = \max_{\theta \in \Theta_t}\norm{B(\theta)}_2$. We define $\set{Z}_t$ as the outer polyhedral approximation to $\set{B}(0,3\sigma_t)$, and $\bar{\set{W}}_t = \set{W} \oplus \bar{B}_t\set{Z}_t$.
\end{itemize}
\noindent
We define the state tube cross sections as the sets: for $k=0,\ldots, N$,
\begin{align}
X_{k|t} = \{x: Tx \leq \alpha_{k|t} \},
\end{align}
where the matrix  $T \in \RR^{d_\alpha \times d_x}$ is chosen such that, for some $\lambda \in [0,1)$, the set $\{x: Tx \leq \mathbf{1} \}$ is $\lambda$-contractive with respect to the system $x_{t+1} = \Phi(\theta)x_t$ for all $\theta\in \Theta_0$. This property is needed to ensure the robust positive invariance of $X_{N|t}$ (see Lemma 5.7 in \cite{kouvaritakis2016model}). To derive the associated linear constraints, we apply a standard result to ensure inclusion of polyhedra (see Proposition 3.31 in \cite{blanchini2008set}). More precisely, for any $j=1,\ldots,m_t$, we have $X_{k|t} \subseteq \{x: \Phi(\theta_t^{(j)}) x + B(\theta_t^{(j)})v_{k|t} + w \in X_{k+1|t}\}$ for all $w\in \bar{\set{W}}_t$ if there exists $H_t^{(j)} \geq 0$ such that:
\begin{subequations}
\begin{align}
&H_t^{(j)}T = T\Phi(\theta_t^{(j)}), \\
&H_t^{(j)}\alpha_{k|t} + TB(\theta_t^{(j)})v_{k|t} + \bar{w}_t\leq \alpha_{k+1|t},
\end{align}
\end{subequations}
where $\bar{w}_t$ is such that $[\bar{w}_t]_i= \max_{w \in \bar{\set{W}}_t}[Tw]_i$ for $i\in \{1,\dots, d_{\alpha}\}$.
\noindent
Similarly, we have $X_{k|t} \subseteq \{x: (F+GK) x + G(v_{k|t} + \zeta)\leq \textbf{1}\}$ for all $\zeta \in \set{Z}_t$, if there exists $H_c \geq 0$, such that:
\begin{subequations}
\begin{align}
&H_c T = F+GK \\
&H_c \alpha_{k|t} + G v_{k|t} + \bar{\zeta}_t \leq \textbf{1},
\end{align}
\end{subequations}
where $\bar{\zeta}_t$ is such that $[\bar{\zeta}_t]_i = \max_{\zeta \in \set{Z}_t} [G\zeta]_i$ for $i\in \{1,\dots, d_c\}$.
\begin{remark}
For simplicity, we have chosen $\bar{w}_t$ and $\bar{\zeta}_t$ so as to be conservative with respect to the noise we apply. However, notice that we could well pre-sample the noise sequence $(\zeta_{t})_{t\ge 0}$ and use it as part of our predictions.
\end{remark}
The matrices $H_t^{(j)}$, and $H_c$ are chosen such that for all $i \in \{ 1,\dots, d_\alpha \}$ and for all $j \in \{1,\dots,m_t\}$,
\begin{subequations}
\begin{align}
\big( H_t^{(j)} \big)_i &= \arg \min_h \{\textbf{1}^\top h: h^\top T = (T)_i \Phi(\theta_t^{(j)}),\ h\geq 0 \},
\label{eq:H^j}
\end{align}
and for all $i \in \{1,\dots, d_c\}$,
\begin{align}
\big( H_c \big)_i &= \arg \min_h \{\textbf{1}^\top h: h^\top T = (F+GK)_i,\ h\geq 0 \}.
\label{eq:H_c}
\end{align}
\end{subequations}
Finally, we have the terminal conditions
\begin{align}
H_t^{(j)} \alpha_N + \bar{w}_t &\leq \alpha_N, \\
H_c\alpha_N + \bar{\zeta}_t &\leq \mathbf{1}.
\end{align}

\subsection{The tube MPC problem}

Fix a step $t$. Let $\boldsymbol{v}_t = \{v_{0|t},\dots,v_{N-1|t}\}$ and $\boldsymbol{\alpha}_t = \{\alpha_{0|t},\dots, \alpha_{N|t}\}$. The resulting tube MPC problem, denoted as $\set{P}_N(x_t,\theta_t,\Theta_t,\bar{\set{W}}_t)$, and evaluated at every $t$, is
\begin{align}
&\underset{\boldsymbol{v}_t,\boldsymbol{\alpha}_t}{\textnormal{minimize}} \sum_{k=0}^{N-1}\left(\ x_{k|t}^\top Q x_{k|t} +  v_{k|t}^\top R v_{k|t}\right) + x_{N|t}^\top P(\theta_t) x_{N|t} \nonumber \\
&\textnormal{subject to}, \textnormal{for all } j=1,\dots,m_t, \textnormal{and } k = 0,\dots,N-1: \nonumber \\
&\textnormal{initial constraints:} \nonumber \\
&\quad x_{0|t} = x_t, \label{eq:initial1} \\
&\quad Tx_{0|t} \leq \alpha_{0|t}, \label{eq:initial2} \\
&\textnormal{system constraints}, \nonumber \\
&\quad x_{k+1|t} = \Phi(\theta_t)x_{k|t} + B(\theta_t)v_{k|t}, \label{eq:system2} \\
&\textnormal{tube constraints}, \nonumber \\
&\quad H_t^{(j)}\alpha_{k|t} + TBv_{k|t} + \bar{w}_t \leq \alpha_{k+1|t}, \label{eq:tube1}\\
&\quad H_c\alpha_{k|t} + G v_{k|t} + \bar{\zeta
}_t \leq \mathbf{1}, \label{eq:tube2}\\
&\textnormal{terminal constraints:} \nonumber \\
&\quad H_t^{(j)}\alpha_{N|t} + \bar{w}_t \leq \alpha_{N|t}, \label{eq:terminal1} \\
&\quad H_c\alpha_{N|t} + \bar{\zeta
}_t\leq \mathbf{1} \label{eq:terminal2}.
\end{align}
where $P(\theta_t)$ is obtained by solving the Lyapunov equation:
\begin{align}
P(\theta_t)-\Phi(\theta_t)^\top P(\theta_t)\Phi(\theta_t) = Q+K^\top R K
\label{eq:lyapunov}
\end{align}
The feasibility of the above problem depends on whether $\theta^\star\in \Theta_t$, i.e., on the event ${\cal G}$. However, for the sake of the analysis, we would like to ensure that the problem the algorithm solves in each step is always feasible (with probability 1). In turn, this will allows us to analyze the performance of the LSE and to establish that indeed ${\cal G}$ holds with probability at least $1-\delta$ (refer to the next section for details). To ensure that the tube MPC problem is always feasible, i.e., even when $\set{G}$ does not occur, we instead solve $\set{P}_N(x_t,\theta_{\rho(t)},\Theta_{\rho(t)},\bar{\set{W}}_t)$\footnote{More precisely, the problem is obtained by replacing $\theta_t$ by $\theta_{\rho(t)}$, $H_t^{(j)}$ by $H_{\rho(t)}^{(j)}$ for all $j$ in $\set{P}_N(x_t,\theta_t,\Theta_t,\bar{\set{W}}_t)$.} in Algorithm \ref{alg:main} where,
\begin{align*}
    \rho(t) := \max\{\tau \leq t: \set{P}_N(x_t,\theta_\tau,\Theta_\tau,\bar{\set{W}}_t) \ \textnormal{is feasible} \}.
\end{align*}
Essentially, in the unlikely event that $\set{P}_N(x_t,\theta_t,\Theta_t,\bar{\set{W}}_t)$ is not feasible, we instead solve $\set{P}_N(x_t,\theta_\tau,\Theta_\tau,\bar{\set{W}}_t)$ using the latest estimates $\theta_\tau$ and $\Theta_\tau$ for which the problem is feasible. This modification is of little practical consequence as we can (and typically do) choose $\delta$ to be very small.
\section{Analysis}\label{sec:analysis}

This section is devoted to the analysis of the Self-tuning Tube-based MPC (STT-MPC) algorithm. We first provides guarantees for the LSE, and then establish the recursive feasibility and the input-to-state stability of the algorithm.

\subsection{Performance of the LSE}

\begin{theorem}
Assume that the optimization problem $\set{P}_N(x_0,\theta_0,\Theta_0,\bar{\set{W}}_0)$ is feasible for initial state $x_0 \in \RR^{d_x}$ and parameter $\theta_0 \in \Theta_0$. Under the STT-MPC algorithm, we have: for all $\delta\in (0,1)$\footnote{Note that the result holds for all $\delta\in (0,1)$ and in particular for the parameter $\delta$ chosen as input of the STT-MPC algorithm.},
$$
\mathbb{P}[{\cal G}]=\mathbb{P}[\|\hat{\theta}_t-\theta^\star\|\le \epsilon_t, \forall t\ge t^\star(\delta)]\ge 1-\delta.
$$
\end{theorem}

\proof The proof follows a similar path as that used to prove Proposition 1 in \cite{jedra2022minimal}, and we provide a sketch of the arguments below. We first decompose the mean square error as (see the proof of Proposition 1 in \cite{jedra2022minimal}):
\begin{align*}
\|\hat{\theta}_t -\theta^\star\|^2 \le  \left\|\left(\sum_{s=0}^{t-2}y_sy_s^\top \right)^{-1/2} \right.&\left.\left(\sum_{s=0}^{t-2}y_sw_s^\top \right)\right\|^2\\
&\times \frac{1}{\lambda_{\min}(\sum_{s=0}^{t-2}y_sy_s^\top)}.
\end{align*}
The first term in the r.h.s. of the above inequality is referred to as the self-normalizing term, and it can be analyzed using Hanson-Wright inequality. Specifically, we can show as in the proof of Proposition 1 in \cite{jedra2022minimal} that it is smaller than $c_4\log(t/\delta)$ with probability at least $1-\delta/2$ for some constant $c_4>0$ and for any $\delta\in (0,1)$. To prove this upper bound, we leverage the fact that the tube MPC problem solved under STT-MPC is feasible in {\it each} step (see Theorem \ref{thm:recursive_feasibility} (ii)), which in turn ensures that $y_t$ is bounded, and simplifies the analysis of the self-normalizing term. Next, we can derive a lower bound of the minimal eigenvalue of the cumulative covariate matrix using the techniques developed in Appendix G in \cite{jedra2022minimal}. As in Proposition 11 in \cite{jedra2022minimal}, we can show that for any $\delta\in (0,1)$, $\lambda_{\min}(\sum_{s=0}^{t-2}y_sy_s^\top)\ge c_5t^{1-\alpha}$ with probability at least $1-\delta$, provided that $t^{1-\alpha}\ge c_6\log(e/\delta)$, for some constants $c_5, c_6>0$. We conclude that for any $\delta\in (0,1)$, $\|\hat{\theta}_t-\theta^\star\|\le \epsilon_t$ with probability at least $1-\delta$ when $t^{1-\alpha}\ge c_6\log(e/\delta)$. The upper bound of $\|\hat{\theta}_t-\theta^\star\|$ can be made uniform over time $t\ge t^\star(\delta)$ applying Lemma 23 in \cite{jedra2022minimal}, which concludes the proof. \ep

\subsection{Recursive feasibility}

\begin{theorem}
If the optimization problem $\set{P}_N(x_0,\theta_0,\Theta_0,\bar{\set{W}}_0)$ is feasible for initial state $x_0 \in \RR^{d_x}$ and parameter $\theta_0 \in \Theta_0$, then, for all $t>0$, 
\begin{enumerate}[(i)]
    \item under event $\set{G}$, the problem $\set{P}_N(x_t,\theta_t,\Theta_t,\bar{\set{W}}_t)$ is feasible;
    \item the problem $\set{P}_N(x_t,\theta_{\rho(t)},\Theta_{\rho(t)},\bar{\set{W}}_t)$ is feasible.
\end{enumerate}
\label{thm:recursive_feasibility} 
\end{theorem}

\proof For (i), we have that $\set{G}$ holds which, together with Assumption \ref{ass:parameter_set}, ensures that $\theta^* \in \Theta_t$ for all $t>0$. Let $\set{P}_N(x_{t-1},\theta_{t-1},\Theta_{t-1},\bar{\set{W}}_{t-1})$ be feasible with solution $(\boldsymbol{v}_{t-1}^*,\boldsymbol{\alpha}_{t-1}^*)$. We define the tail point $(\hat{\boldsymbol{v}}_t,\hat{\boldsymbol{\alpha}}_t)$ as
\begin{align*}
    &\hat{\boldsymbol{v}}_t = \{v^*_{1|t-1},\dots,v^*_{N-1|t-1},0\}, \\ &\hat{\boldsymbol{\alpha}}_t = \{\alpha_{1|t-1},\dots,\alpha_{N|t-1},\alpha_{N|t-1}\},
\end{align*}
and we show that it is a feasible point for $\set{P}_N(x_{t},\theta_{t},\Theta_{t},\bar{\set{W}}_{t})$, i.e., that it satisfies the constraints \eqref{eq:initial2}-\eqref{eq:terminal2} at time $t$.

The initial constraint \eqref{eq:initial2} is satisfied since $(\boldsymbol{v}_{t-1}^*,\boldsymbol{\alpha}_{t-1}^*)$ satisfies \eqref{eq:tube1}, and $\theta^* \in \Theta_{t-1}$, which together imply that $x_t \in \set{X}_{1|t-1}$. The system dynamics constraint \eqref{eq:system2} holds automatically. We now consider only the vertices of $\Theta_{t-1}$ and note that $\bar{w}_{t-1} \geq \bar{w}_t$ and $\bar{\zeta}_{t-1} \geq \bar{\zeta}_t$. At time $t$, the tube constraints \eqref{eq:tube1} and \eqref{eq:tube2} for $k \in \{0,\dots,N-2 \}$ are satisfied whenever the tube constraints for $k \in \{1,\dots,N-1 \}$ at time $t-1$ are satisfied. Furthermore, noting that $v_{N-1|t} = 0$ and $\alpha_{N|t} = \alpha^*_{N|t-1}$, by simple substitution, we will have that \eqref{eq:tube1} and \eqref{eq:tube2} for $k = N-1$ at time $t$ hold when the terminal constraints \eqref{eq:terminal1} and \eqref{eq:terminal2} at time $t-1$ hold. If the terminal constraints hold at time $t-1$, they will also hold at time $t$. Finally, since $\Theta_t \subseteq \Theta_{t-1}$, these constraints will also hold for every vertex of $\Theta_t$. 

For (ii) note that, due to Assumption \ref{ass:parameter_set}, $\theta^* \in \Theta_0$ so that there will always exist $\tau \leq t$ such that $\theta^* \in \Theta_\tau$ and such that $\set{P}_N(x_{\tau},\theta_{\tau},\Theta_{\tau},\bar{\set{W}}_{\tau})$ is feasible. Then, using a similar argument as in (i) we will have that $\set{P}_N(x_{\tau+1},\theta_{\tau},\Theta_{\tau},\bar{\set{W}}_{\tau+1})$ is feasible and thus $\set{P}_N(x_{t+1},\theta_{\tau},\Theta_{\tau},\bar{\set{W}}_{t+1})$ is feasible.
\ep

\subsection{Input-to-State Stability (ISS)}

Let $X_{\set{P}}$ be the set of states $x_0$ such that problem $\set{P}_N(x_{0},\theta_{0},\Theta_{0},\bar{\set{W}}_{0})$ is feasible. Additionally, let $J(x_t,\boldsymbol{v}_t,\theta_t)$ be the objective of problem $\set{P}_N(x_{t},\theta_{t},\Theta_{t},\bar{\set{W}}_{t})$, let $\boldsymbol{v}_t^*$ be its optimal solution and let $V^*(x_t,\theta_t)$ be its associated value.
To simplify notations, we refer to the closed loop system \eqref{eq:newdyn} as,
\begin{align}
x_{t+1} = f(x_t,v_t,w_t,\theta_t)
\label{eq:linear_map}
\end{align}
where, we refer to $B(\theta_t)\zeta_t+w_t$ as $w_t$. We further define the stage cost $L(x,v) = x^\top( Q+K^\top R K) x  + v^\top R v$, and terminal cost $\phi(x,\theta) = x^\top P(\theta) x$ so that
\begin{align*}
    J(x_t,\boldsymbol{v}_t,\theta_t) = \sum_{k=0}^{N-1}L(x_{k|t},v_{k|t}) + \phi(x_{N|t},\theta_t).
\end{align*}

\begin{lemma} [Theorem 3 in \cite{limon2009input}]
The system \eqref{eq:linear_map} with control law $v_t(x_t,\theta_t)$ is ISS with region of attraction $\set{R} \subseteq \RR^{d_x}$ if
\begin{enumerate}[(i)]
\item $\set{R}$ is a compact set containing the origin, and is robust positive invariant for \eqref{eq:linear_map}.
\item There exist $\set{K}_\infty$-functions $\alpha_1, \alpha_2, \alpha_3$, $\set{K}$-functions $\sigma_1, \sigma_2$, and a continuous Lyapunov function $V: \set{R} \to \RR_+$ such that for all states $x_t \in \set{R}$ and disturbances $w_t \in \bar{\set{W}}_t$ we have,
\begin{align}
\label{eq:lyapunov-bound}
\alpha_1(\norm{x_t}) \leq V(x_t) &\leq \alpha_2(\norm{x_t})&\\
V(x_{t+1}) - V(x_t) &\leq -\alpha_3(\norm{x_t}) +\sigma_1(\norm{w_t}) \nonumber\\ 
& \quad + \sigma_2(\norm{\theta_t - \theta^*}). 
\label{eq:lyapunov_difference_bound}
\end{align} 
\end{enumerate} 
\label{lem:ISS-Lyapunov}
\end{lemma}
\begin{theorem}
Under event $\set{G}$ and under the STT-MPC algorithm, for all initial conditions $x_0 \in X_{\set{P}}$ and all parameters $\theta^* \in \Theta_0$, the system \eqref{eq:linear_system} is ISS with region of attraction $X_{\set{P}}$.
\label{thm:ISS}	
\end{theorem}

It suffices to show that $X_{\set{P}}$ satisfies condition (i) of Lemma \ref{lem:ISS-Lyapunov} and $V^*(x_t,\theta_t)$ is a Lyapunov function satisfying condition (ii).

For condition (i), suppose that $\set{P}_N(x_{0},\theta_{0},\Theta_{0},\bar{\set{W}}_{0})$ is feasible, i.e. $X_{\set{P}}$ is not empty. Then the terminal constraints \eqref{eq:terminal1} and \eqref{eq:terminal2} ensure that there exists an $\alpha_{N|0}$ such that the terminal cross-section set $X_{N|0} = \{x: Tx \leq \alpha_{N|0} \}$ satisfies the constraints \eqref{eq:constraints} and $f(x,0,\theta,w) \in X_{N|0}$ for all $x\in X_{N|0}$, $w \in \bar{\set{W}}_0$, and $\theta \in \Theta_0$, i.e. it is robustly invariant. For every $x \in X_{N|0}$, the point $(\boldsymbol{v},\boldsymbol{\alpha}) = (\{0,\dots, 0 \},\{\alpha_{N|0}, \dots, \alpha_{N|0} \})$ is feasible for $\set{P}_N(x,\theta_{0},\Theta_{0},\bar{\set{W}}_{0})$, thus $X_{N|0} \subseteq X_{\set{P}}$. The robust positive invariance of $X_{\set{P}}$ will then follow directly from Theorem \ref{thm:recursive_feasibility}. Since $X_{N|t} \subseteq X_{\set{P}}$ and $0 \in \bar{\set{W}}_t$ by Assumption \ref{ass:noise}, we have $0 \in X_{\set{P}}$. Finally, $X_{\set{P}}$ is compact due to Assumption \ref{ass:constraints}.

For condition (ii) we first show that the bound \eqref{eq:lyapunov-bound} holds for $V^\star(x_t,\theta_t)$. For any given state $x_t$, and parameters $\theta_t \in \Theta_t$, and for any $Q, R \succ 0$, problem $\set{P}_N(x_t,\theta_{t},\Theta_{t},\bar{\set{W}}_{t})$ will be a quadratic program and it follows from Theorem 4 in \cite{bemporad2002explicit} that $V^\star(x_t,\theta_t)$ is a continuous function of $x_t$. Since $\Theta_t$ is compact by Assumption \ref{ass:parameter_set}, it follows from Lemma 4.2 in \cite{khalil2015nonlinear} that there exist $\set{K}_\infty$-functions $\alpha_1$, $\alpha_2$ such that for any $x_t \in X_{\set{P}}$,
\begin{align*}
\alpha_1(\norm{x_t}) \leq V^\star(x_t,\theta_t) \leq \alpha_2(\norm{x_t}).
\end{align*}
We now show the bound \eqref{eq:lyapunov_difference_bound}. First, note that the continuity of $f(x,u,w,\theta)$, $L(x,u)$, and $\phi(x,\theta)$ with respect to $x,u,w$, and $\theta$, implies that there exist $\set{K}_\infty$-functions $\sigma_x, \sigma_w, \sigma_\theta, \sigma_L, \sigma_\phi$, so that for all $x,x_1,x_2 \in X_{\set{P}}$, $v \in \RR^{d_u}$, $w,w_1,w_2 \in \bar{\set{W}}_0$, and $\theta, \theta_1, \theta_2 \in \Theta_t$,
\begin{align*}
\norm{f(x_1,v,w_1,\theta_1) - f(x_2,v,w_2,\theta_2)} &\leq \sigma_x(\norm{x_1 - x_2}) \\
 +\sigma_w(\norm{w_1 - w_2}) &+ \sigma_\theta(\norm{\theta_1 - \theta_2}), \\
| L(x_1,v) - L(x_2,v)| &\leq \sigma_L(\norm{x_1-x_2}),\\
\norm{\phi(x_1,\theta) - \phi(x_2,\theta)} &\leq \sigma_\phi(\norm{x_1 - x_2}).
\end{align*} 
Recall that $\boldsymbol{v}_t^* = \{v_{0|t}^*,\dots, v_{N-1|t}^* \}$ is the predicted control sequence that arises from the solution of $\set{P}_N(x_t,\theta_{t},\Theta_{t},\bar{\set{W}}_{t})$. We denote by $\boldsymbol{x}_t^* = \{x_{0|t}^*, \dots, x_{N|t}^*\}$, the associated predicted state trajectory (using \eqref{eq:prediction_dynamics}, with parameter $\theta_t$, when applying $\boldsymbol{v}_t^*$). Similarly, recall the tail control sequence $\hat{\boldsymbol{v}}_{t+1}:= \{\hat{v}_{0|t+1},\dots, \hat{v}_{N-1|t+1} \} = \{ v_{1|t}^*,\dots, v_{N-1|t}^*, 0 \}$ for which we define, an associated predicted state trajectory $\hat{\boldsymbol{x}}_{t+1} = \{ \hat{x}_{0|t+1}, \dots, \hat{x}_{N|t+1}\}$ (using \eqref{eq:prediction_dynamics} with parameter $\theta_{t+1}$, when applying $\hat{\boldsymbol{v}}_{t+1}$) with the property that $\hat{x}_{0|t+1} = x_{1|t}^*$. To simplify the presentation, we omit references to the subscript $t$ when there is no ambiguity. Noting the sub-optimality of $\hat{\boldsymbol{v}}_{t+1}$, we will have
\begin{align*}
    J(&x_{t+1},\hat{\boldsymbol{v}}_{t+1},\theta_{t+1}) - J(x_{t},\boldsymbol{v}^*_{t},\theta_{t}) \\
    &\leq V^*(x_{t+1},\theta_{t+1}) - V^*(x_t,\theta_t) \\
    & \leq \sum_{k=0}^{N-1}L(\hat{x}_{k},\hat{v}_{k}) - L(x_{k}^*,v_{k}^*) + \phi(\hat{x}_{N},\theta_{t+1}) - \phi(x_{N}^*,\theta_t) \\
    & =-L(x_{0|t},v^*_{0}) +\sum_{k=0}^{N-2} L(\hat{x}_{k},v_{k+1}^*) - L(x_{k+1}^*,v_{k+1}^*) \\
    & \quad + L(\hat{x}_{N-1},0) + \phi(\hat{x}_{N},\theta_{t+1}) - \phi(\hat{x}_{N-1},\theta_{t+1}) \\
    & \quad + \phi(\hat{x}_{N-1},\theta_{t+1}) - \phi(x_{N}^*,\theta_t).
\end{align*}
We will proceed to upper bound the above quantities. We first note that, for some $\set{K}_\infty$ function $\kappa_\epsilon$ we will have,
\begin{align*}
\sigma_\theta (\norm{\theta_{t+1} - \theta_t}) \leq \sigma_\theta (\norm{\theta_{t} - \theta^*}) + \kappa_\epsilon(\epsilon_{t})
\end{align*}
and for all $k\in \{1,\dots,N-1\}$,
\begin{align*}
    \norm{\hat{x}_{k} - x^*_{k+1}} &\leq \sigma_x(\norm{\hat{x}_{k-1} - x^*_{k}}) + \sigma_\theta(\norm{\theta_{t+1} - \theta_t}) \\
    &\leq \sigma_x(\norm{\hat{x}_{k-1} - x^*_{k}}) + \sigma_\theta(\norm{\theta_{t} - \theta^*}) + \kappa_\epsilon(\epsilon_{t}).
\end{align*}
Here we the triangle inequality, the fact that event $\set{G}$ holds, and the monotonicity of $\epsilon_t$. Applying this relation recursively leads to,
\begin{align*}
    \norm{\hat{x}_{k} - x^*_{k+1}} \leq &\frac{1}{2}(2\sigma_x)^k(\norm{\hat{x}_{0} - x^*_{1}}) \\ &+ \sum_{i=0}^{k-1}\frac{1}{2}(2\sigma_x)^i \circ 2(\sigma_\theta (\norm{\theta_{t} - \theta^*} + \kappa_\epsilon(\epsilon_{t})),
\end{align*}
where $\norm{\hat{x}_{0} - x^*_{1}} \leq \sigma_w(w_t) + \sigma_\theta (\norm{\theta_{t} - \theta^*}) + \kappa_\epsilon(\epsilon_{t})$.

Now, since $Q, R \geq 0$, there exists a $K_\infty$-function $\alpha_3$ such that $-L(x_{0|t},v^*_{0}) \leq -\alpha_3(\norm{x_{0|t}})$. We will also have, for all $k \in \{0,\dots ,N-2\}$,
\begin{align*}
    |L(\hat{x}_{k},v_{k+1}^*) - L(x_{k+1}^*,v_{k+1}^*)| \leq \sigma_L(\norm{\hat{x}_{k} - x^*_{k+1}}).
\end{align*}
Then, from \eqref{eq:lyapunov} it follows that,
\begin{align*}
    L(\hat{x}_{N-1},0) + \phi(\hat{x}_{N,\theta_{t+1}}) - \phi(\hat{x}_{N-1},\theta_{t+1}) = 0.
\end{align*}
Finally, recall that \eqref{eq:lyapunov} is linear in $P(\theta)$ and note that its solution is unique for all $\theta \in \Theta_0$ (Theorem 3.7 \cite{khalil2015nonlinear} using that $\Phi(\theta)$ is stable). It follows then that $P(\theta)$ is Lipschitz continuous and that there will exist a $\set{K}_\infty$-function $\kappa_\phi$ such that,
\begin{align*}
    |\phi(\hat{x}_{N-1},\theta_{t+1}) - \phi(x^*_{N},\theta_{t})| &\leq \sigma_\phi(\norm{\hat{x}_{N-1} - x^*_{N}}) \\ &+ \kappa_\phi(\norm{\theta_{t} - \theta^*}) + \kappa_\epsilon(\epsilon_{t}).
\end{align*}
We now group the above terms together to get,
\begin{align*} V^*(x_{t+1},\theta_{t+1}) - V^*(x_t,\theta_t) \leq& -\alpha_3(\norm{x_{0|t}}) + \bar{\sigma}_w(\norm{w_t}) \\
    &+ \bar{\sigma}_\theta(\norm{\theta_t - \theta^*}) + \bar{\kappa}_\epsilon(e_t)
\end{align*}
where $\bar{\sigma}_w, \bar{\sigma}_\theta$, and $\bar{\kappa}_\epsilon$ are $\set{K}$-functions. Since $e_t \to 0$ as $t\to \infty$, we will have that condition (iii) is satisfied.
\ep

\section{Numerical Experiments} \label{sec:numerical}
\begin{figure}[t]
    \centering
    \includegraphics[width=\linewidth]{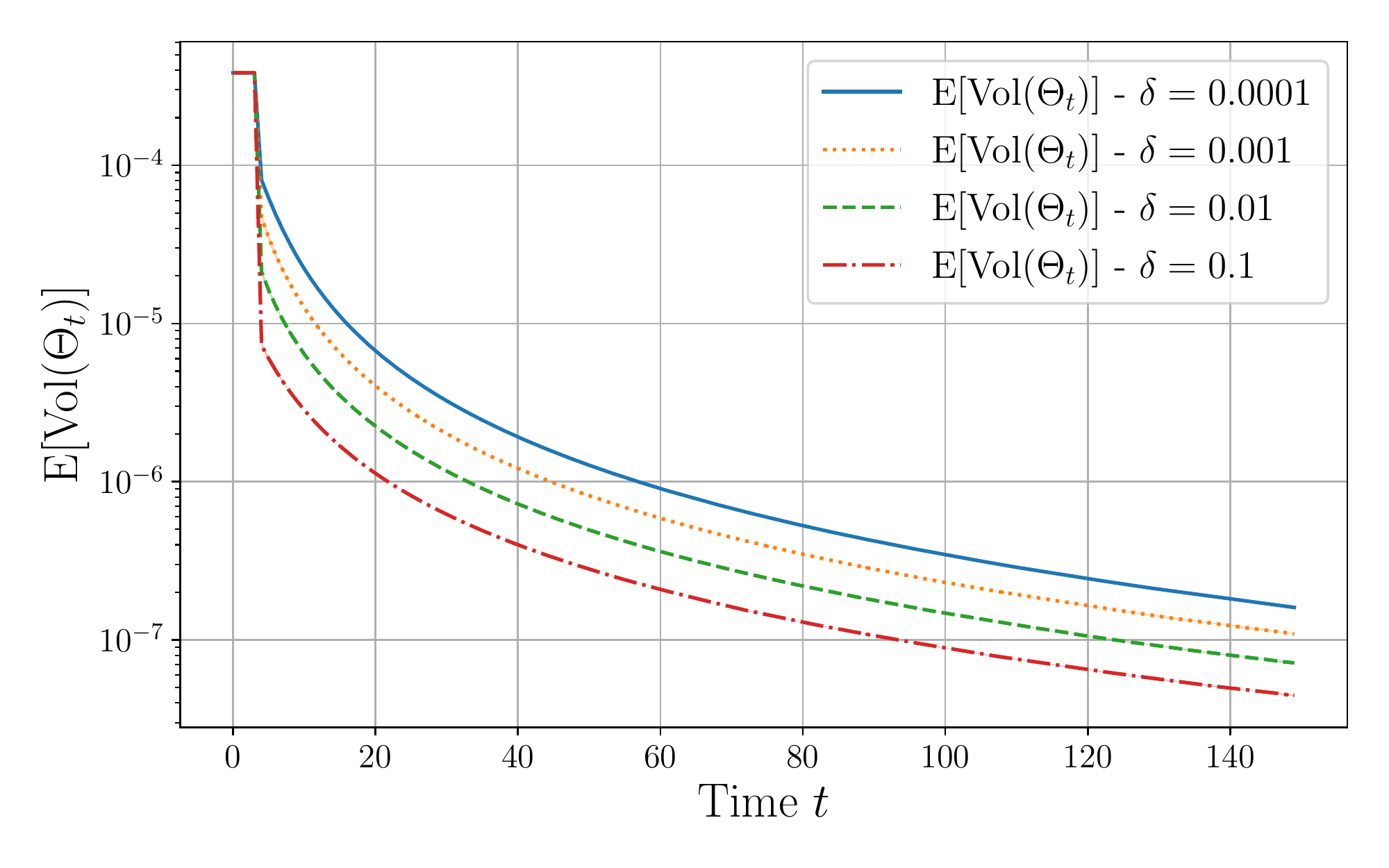}
    \caption{Average volume of $\textrm{Vol}(\Theta_t)$ for different values of $\delta$. Results are averaged over 10 different runs (the small shadowed areas depicts 99.99\% confidence interval).}
    \label{fig:volume_shrinking}
\end{figure}
We illustrate the performance of STT-MPC on a second-order linear systems with true parameters
\begin{equation}
    A = \begin{bmatrix}
    0.6 & 0.2\\-0.1 & 0.4
    \end{bmatrix},\quad
    B = \begin{bmatrix}
    1\\0.6
    \end{bmatrix},
    \quad \sigma=0.01,
\end{equation}
where $B$ is assumed to be known, while $A$ is the uncertain parameter. The initial state is $x_0=(6,3)$, and we consider $\Theta_0$ to be a $4$-dimensional hypercube centered at $\theta_0=\begin{bmatrix}0.57&0.17 &-0.12 & 0.42\end{bmatrix} $ with side length $0.14$. Consequently, the resulting stabilizing feedback gain is $K=\begin{bmatrix} -0.426 & -0.290\end{bmatrix}$. The least-squares confidence region $\Delta_t$ is over-approximated by an hypercube centered in $\theta_t$ with side length $4\varepsilon_t$. We consider the following state and input constraints: $[x_t]_1 \geq -0.15, [x_t]_2 \geq -1.1$ and $u_t \leq 0.5$.
 The matrix $T$ is computed according to the relation (5.98) in \cite{kouvaritakis2016model}, with $\lambda=0.999$. We consider $Q=I_{d_x}$, $R=I_{d_u}$, and, for simplicity, we let the worst case noise realization be $\bar w_t = \bar w_0$, for all $t\geq 1$.
All the simulations were performed in Python 3.9\footnote{Find  the code here: \url{https://github.com/rssalessio/SelfTuningTubeBasedMPC}}, using the CVXPY library \cite{diamond2016cvxpy} and the MOSEK solver.

\begin{figure}[t]
    \centering
    \includegraphics[width=\linewidth]{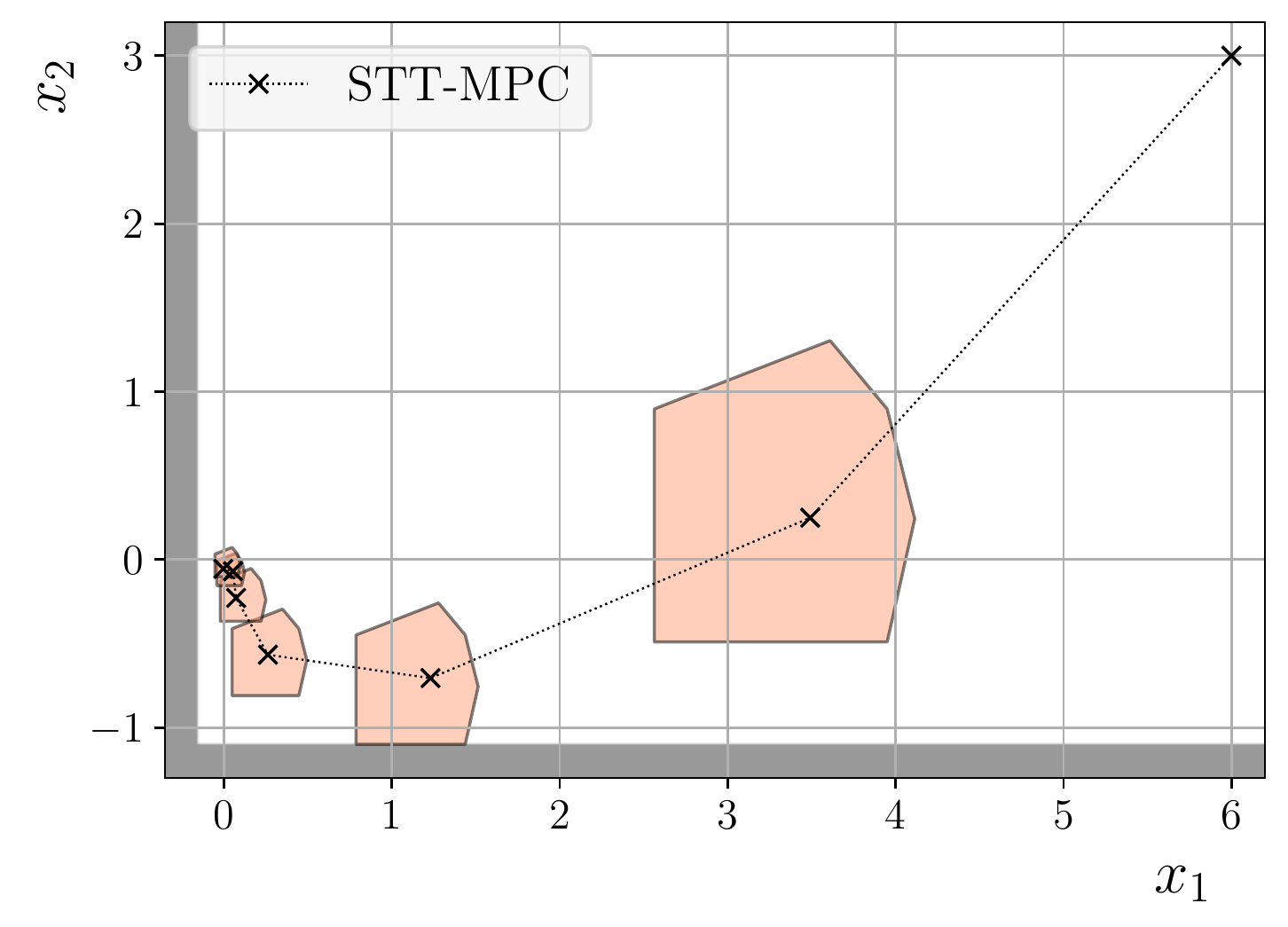}
    \caption{Illustration of the state trajectory using \textsc{STT-MPC}, and its tubes, over a single run. The gray areas depict the state constraints  $[x_t]_1\geq -0.15, [x_t]_2\geq -1.1$. As a comparison, the plot of the state trajectory under $u_t=Kx_t$ is also shown.}
    \label{fig:control}
\end{figure}

\begin{table}[!ht]
\centering
\label{table:volume_reduction}
\caption{Average volume of $\textrm{Vol}(\Theta_t)$ with respect to $\textrm{Vol}(\Theta_0)$}
\begin{tabular}{@{}llllll@{}}
\toprule
Time step $t$ &           1&  5&  15 & 50 & 100\\\toprule
$\delta = 10^{-1}$     &  $100\%$ &  $1.312\%$ &  $0.401\%$ & $0.071\%$ & $0.023\%$\\
$\delta = 10^{-2}$     &  $100\%$ &  $3.354\%$ &  $0.827\%$ & $0.125\%$ & $0.038\%$\\
$\delta = 10^{-3}$     &  $100\%$ &  $7.058\%$ &  $1.525\%$ & $0.205\%$ & $0.059\%$\\
$\delta = 10^{-4}$     &  $100\%$ &  $12.651\%$ &  $2.593\%$ & $0.319\%$ & $0.088\%$\\\bottomrule
\end{tabular}
\end{table}
Figure \ref{fig:volume_shrinking} shows the average volume of $\Theta_t$ over 10 different runs. We note that the reduction in volume is significant, and well below $1\%$ of the volume of $\Theta_0$ after $50$ steps (see also the details in table \ref{table:volume_reduction}; after 100 steps the volume is reduced below $0.1\%$ of the original volume).
In Figure \ref{fig:control} we present the trajectory of a single run of \textsc{STT-MPC}. In the plot are shown the tubes computed by the algorithm. We note that the constraints are robustly satisfied. Moreover, we also note a sensible reduction of the tubes over time.  
\section{Conclusions} \label{sec:conclusion}

We proposed STT-MPC which combines least-squares estimaton with a polytopic tube-based MPC method to ensure robust constraint satisfaction while learning the system dynamics. In contrast with existing conservative set-based identification approaches, we construct parameter sets that hold for all time with a confidence level $\delta$ chosen as a design parameter. Persistent excitation is ensured by injecting a truncated noise signal which decays at a rate $t^{-\alpha}$, with $\alpha > 0$ controlling the trade-off between (transient) estimation accuracy and controller performance. Importantly, we asymptotically recover the performance of the equivalent tube-based MPC that has full knowledge of the dynamics.

We analysed our proposed algorithm by deriving performance guarantees for the LSE, as well as by establishing recursive feasibility and input-to-state stability guarantees. We demonstrated the performance of our adaptation scheme, its dependence on $\delta$, and the robust constraint satisfaction of STT-MPC via a numerical example. 

Future work will focus on the analysis of the performance of our algorithm and quantifying the additional objective cost that must be paid for learning the dynamics. In particular, we will investigate whether a PE noise decaying as $1/\sqrt{t}$ could ensure a {\it regret} scaling at most as $\sqrt{t}$ (as this was established in \cite{jedra2022minimal} for LQR problems). 

\bibliography{RL.bib}
\bibliographystyle{IEEEtran}

\end{document}